\documentclass[a4paper,11pt]{article}
\usepackage{pos}
\usepackage{siunitx}
\usepackage{wrapfig}
\usepackage{caption}
\usepackage{subcaption}
\usepackage{braket}
\usepackage{amsmath}
\usepackage[rightcaption]{sidecap}

\usepackage{ragged2e}
\usepackage{etoolbox}
\apptocmd{\thebibliography}{\justifying}{}{}

\title{An accelerator-based source of high-intensity quantum-entangled annihilation gamma photons}

\ShortTitle{High-intensity quantum-entangled gamma source}

\author*[a]{Riad Suleiman}

\affiliation[a]{Thomas Jefferson National Accelerator Facility, \\ Newport News, Virginia 23606, USA}

\emailAdd{suleiman@jlab.org}

\abstract{We present the concept and development of a novel accelerator-based source of high-intensity quantum-entangled 511 keV gamma-ray pairs produced through positron-electron annihilation. 
The source leverages the unique capabilities of the proposed Jefferson Lab positron facility to generate polarized, high-current positron beams with a well-defined time structure.
These beams enable the production of entangled annihilation photons at intensities far exceeding those available from conventional radioactive sources.
The resulting gamma-ray pairs can be characterized using Compton polarimetry techniques, providing a powerful platform for precision studies of quantum entanglement. 
The combination of high intensity, controllable polarization, and precise timing also opens new opportunities in medical imaging, materials science, quantum information science, and spintronics.
Compared with traditional radioactive sources, the proposed system offers orders-of-magnitude higher photon flux, tunable beam parameters, and unprecedented control of the annihilation process. 
We discuss the source design, methods for entanglement characterization, potential applications, and future development directions.}

\FullConference{International Workshop on Low Energy Electron Positron Physics at Jefferson Lab (LEEPP2026)\\
23--27 March 2026\\
Thomas Jefferson National Accelerator Facility, Virginia, USA\\}

\begin{document}
\maketitle

\section{Introduction}

Quantum entanglement is one of the most fundamental and intriguing features of quantum mechanics, with far-reaching implications for both basic science and emerging quantum technologies. 
To date, most experimental studies of entangled photon pairs have relied on optical systems operating at photon energies in the eV range. 
In contrast, entangled gamma rays produced through positron–electron annihilation offer a unique opportunity to investigate quantum correlations in the high-energy regime. 
When a positron annihilates with an electron, the interaction produces two \SI{511}{keV} gamma photons emitted in opposite directions. 
Conservation laws require the two-photon system to be generated in an entangled polarization state. 
Although such entangled gamma pairs have previously been studied using radioactive sources~\cite{Ujeniuc:2024hse}, accelerator-based positron beams provide a highly controlled and scalable alternative, enabling substantially higher intensities, precise timing structures, and tunable beam parameters. 

Positrons are produced in the decay of several radioactive isotopes, with the most common source being the $\beta^+$ decay of $^{22}$Na $(p \rightarrow n + e^+ + \nu_e)$. Conventional radioactive positron sources are typically limited to activities of about \SI{1}{GBq} (\SI{27}{mCi}), corresponding to approximately $10^9$ annihilations per second. 
In this work, we present a novel approach for generating quantum-entangled gamma photons using an accelerator-driven positron beam. 
This technique provides unprecedented control over the parameters of the positron beam, including the intensity, polarization, energy, and time structure, enabling systematic investigations of quantum entanglement and opening new opportunities for a wide range of applications.

\section{Accelerator-Based Positron Source}

\begin{table}[th!]
\caption{Key parameters of the positron beam expected at Jefferson Lab, including beam intensity, polarization, energy, bunch structure, and timing characteristics relevant for the production of entangled \SI{511}{keV} gamma photons.}
\centering
\begin{tabular}{l|l}
\hline \hline
Parameter & Value \\ \hline 
Beam Repetition Rate & \SIrange{1}{1500}{\mega\hertz} \\
Duty Factor & 100\% \\
Unpolarized Positron Intensity & > \SI{1}{\micro\ampere} \\
Polarized Positron Intensity & > \SI{50}{nA} \\
Beam Energy & \SIrange{1}{10}{\MeV} \\
Beam Polarization, any direction & $> 60\%$ \\
Relative Momentum Spread, $\sigma_{(\Delta p/p)}$ & 
$< 1\%$ \\
Bunch Length, $\sigma_z$ & $< \SI{4}{\ps}$ \\
Beam Emittances, $\epsilon_{nx}$ and $\epsilon_{ny}$ &
$< \SI{e-3}{\meter\radian}$\\
Polarization Reversal Rate & \SI{5}{\kHz} \\ \hline \hline
\end{tabular}
\label{tab:PosBeam_pars}
\end{table}

Jefferson Lab provides a unique environment for producing intense positron beams. The facility will be capable of delivering unpolarized positron beams that exceed $6 \times 10^{12}$ positrons per second, polarized positron beams above $3 \times 10^{11}$ positrons per second, and precisely controlled time structures.
Table~\ref{tab:PosBeam_pars} summarizes the expected properties of the positron beam~\cite{eric_voutier,habet:ipac22-mopotk012}. 
 
\section{Quantum-Entangled Gamma Source Concept}
 
\begin{figure}[h]
  \centering
  \begin{minipage}[c]{0.6\textwidth} 
    \includegraphics[width=\textwidth]{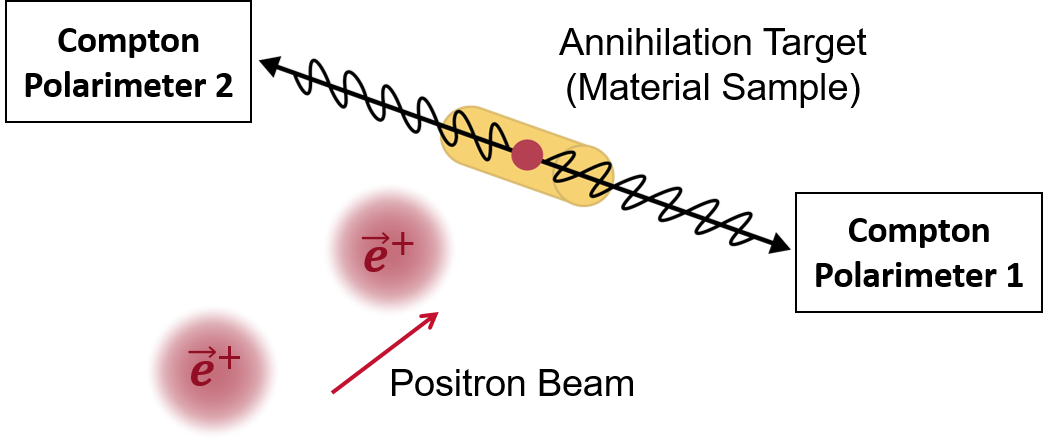}
  \end{minipage}%
  \hfill
  \begin{minipage}[c]{0.35\textwidth} 
    \caption{Illustration of the entangled gamma source driven by a well-defined positron beam. 
    The positron bunch spacing can be varied from sub-ns to many \textmu s, providing a flexible timing structure for coincidence and lifetime measurements.}
    \label{fig:fig1}
  \end{minipage}
\end{figure}

The proposed entangled gamma-ray source, illustrated in Fig.~\ref{fig:fig1}, consists of four principal components. First, a high-intensity positron beam is delivered by the accelerator with a well-defined energy, polarization, and time structure. 
Second, the positrons are stopped in an annihilation target, such as a metallic foil or other suitable material, where they annihilate with electrons in the target. 
Third, the annihilation process produces 
pairs of \SI{511}{keV} gamma photons emitted in opposite directions and exhibiting quantum-entangled polarization states. 
Finally, the resulting gamma pairs are analyzed using Compton polarimeters comprising high-resolution gamma-ray detectors, coincidence electronics, and a data acquisition system capable of recording correlated events with high timing and energy precision.
  
\section{Compton Polarimetry and Entanglement Measurement}

\begin{figure}[h]
  \centering
  \begin{minipage}[c]{0.65\textwidth} 
    \includegraphics[width=\textwidth]{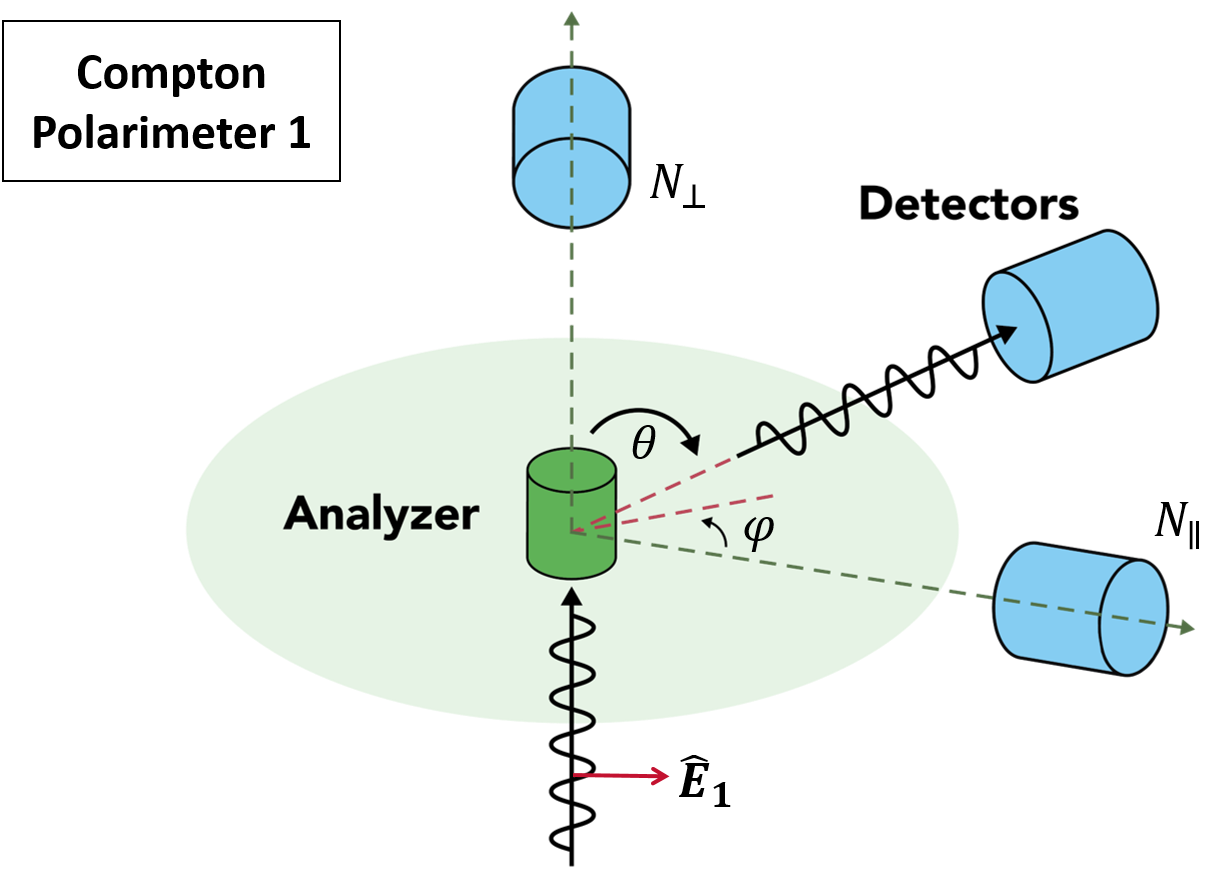}
  \end{minipage}%
  \hfill
  \begin{minipage}[c]{0.3\textwidth} 
    \captionof{figure}{Layout of a single Compton polarimeter showing three detectors arranged around a central analyzer, illustrating a possible configuration for full azimuthal-angle coverage.
    The incident gamma, with polarization vector $\hat{E}_1$, undergoes Compton scattering in the analyzer. The detectors $N_\parallel$ and $N_\perp$ measure scattered gammas parallel and perpendicular, respectively, to the plane defined by the propagation direction and polarization vector of the incident gamma.}
    \label{fig:fig2}
  \end{minipage}
\end{figure}

The linear polarization of the gamma photons is measured using Compton polarimeters. Figure~\ref{fig:fig2} shows the Compton polarimeter layout used for polarization analysis.

\subsection{Single Compton Polarimeter}

\begin{figure}
     \centering
     \begin{subfigure}[b]{0.49\textwidth}
         \centering
         \includegraphics[width=\textwidth]{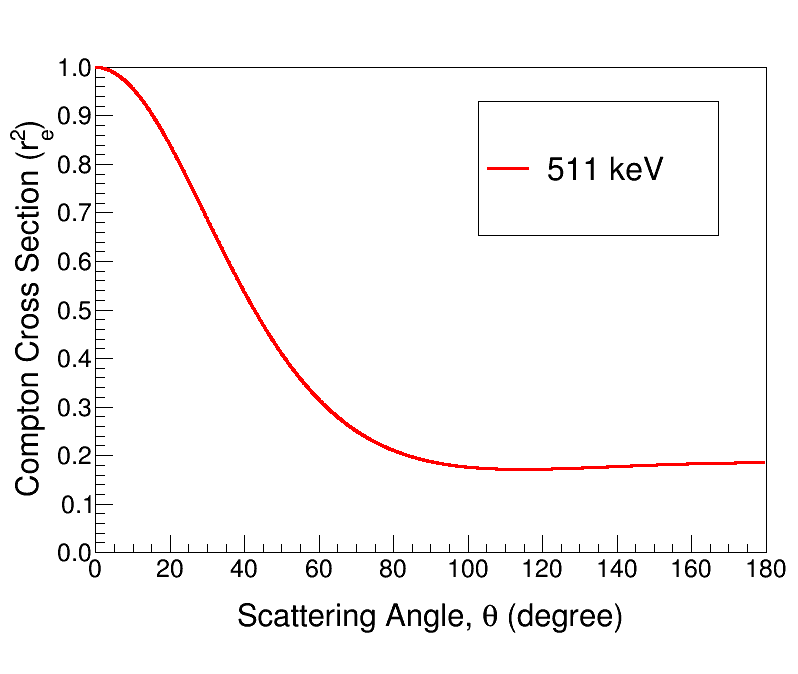}
     \end{subfigure}
     \hfill
     \begin{subfigure}[b]{0.49\textwidth}
         \centering
         \includegraphics[width=\textwidth]{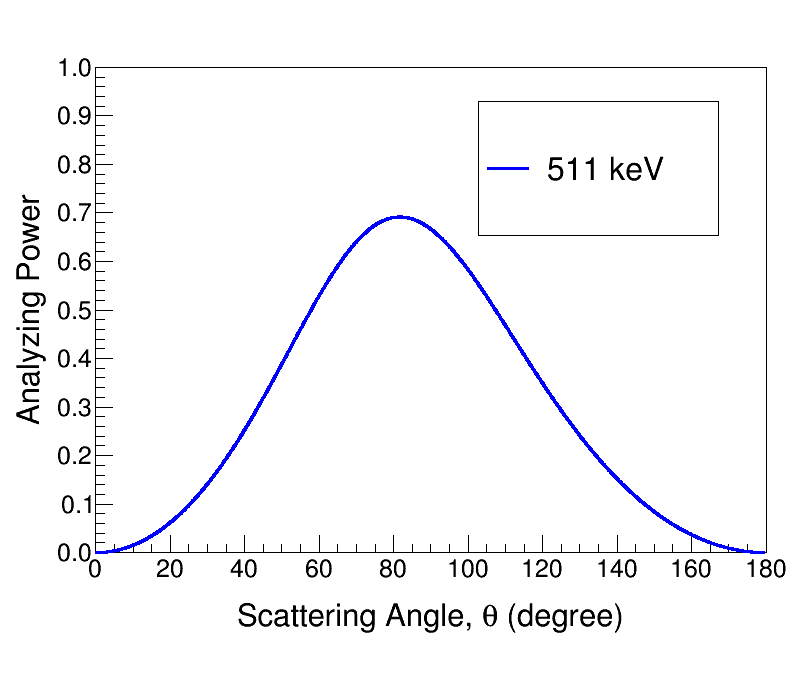}
     \end{subfigure}
        \caption{The unpolarized differential Compton scattering cross section (left) and the corresponding Compton-scattering analyzing power (right) for 511 keV gamma rays. Here, $r_e = 2.82\times10^{-15}$ m denotes the classical electron radius.}
        \label{fig:CS_AP_graphs}
\end{figure}

The polarization of gamma photons can be analyzed via Compton scattering. The differential cross section depends on the polarization state of the incoming gamma, providing sensitivity to its polarization orientation.
The analyzing power $A_p$ is large at specific scattering angles, allowing optimized detector configurations. For \SI{511}{keV} gamma photons, the unpolarized differential Compton scattering cross section and  the Compton scattering analyzing power are shown in Fig.~\ref{fig:CS_AP_graphs}.

\subsection{Double Compton Polarimeters}

\begin{figure}[h]
  \centering
  \begin{minipage}[c]{0.65\textwidth} 
    \includegraphics[width=\textwidth]{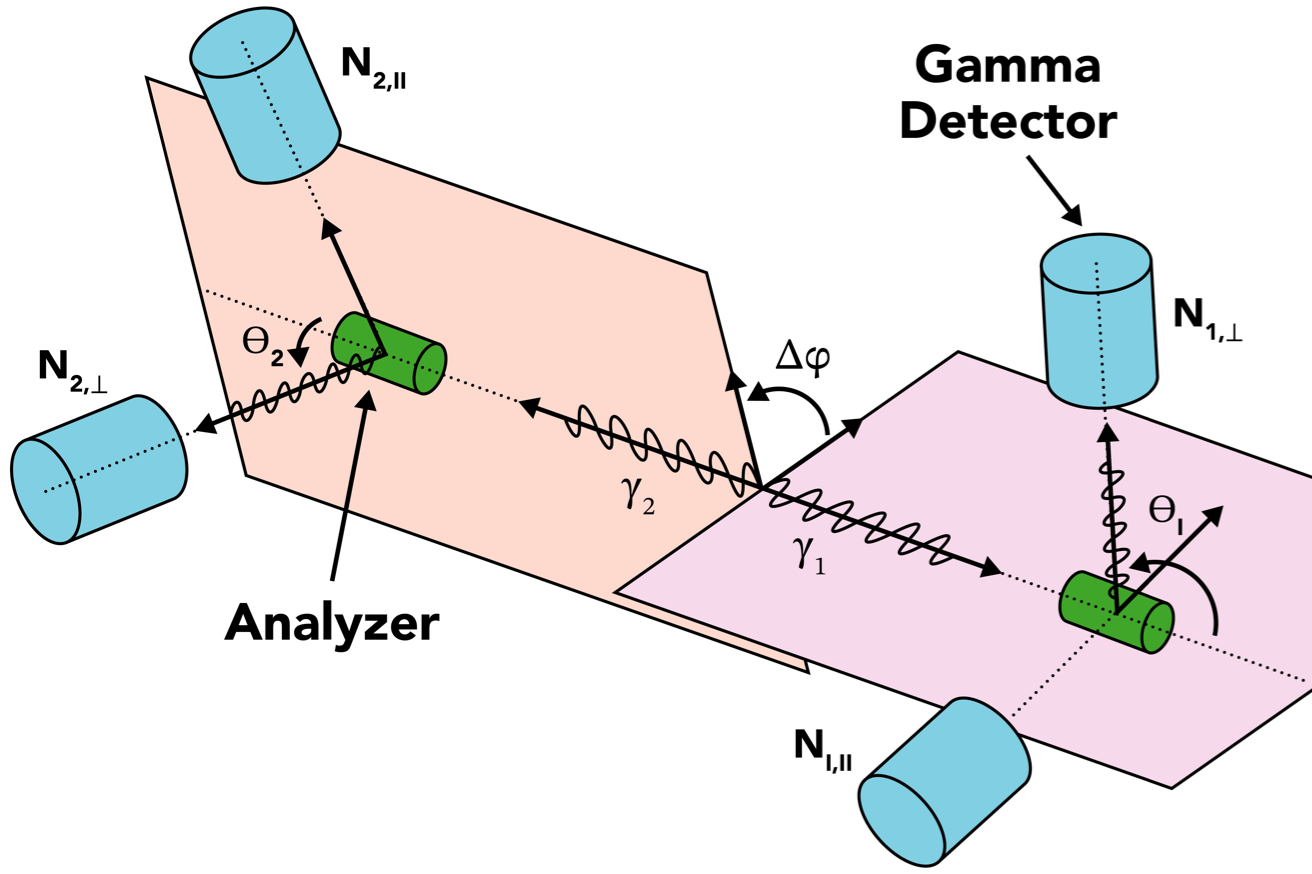}
  \end{minipage}%
  \hfill
  \begin{minipage}[c]{0.3\textwidth} 
    \caption{Schematic layout of the double Compton polarimeter. 
    Entangled \SI{511}{keV} gamma photons are detected by two opposing Compton polarimeters, enabling measurement of polarization dependent scattering correlations and tests of quantum entanglement.}
    \label{fig:fig4}
  \end{minipage}
\end{figure}

To probe entanglement, both photons in the annihilation pair are measured using separate Compton polarimeters~\cite{PhysRevA.111.053708}. The double differential cross section encodes correlations between the polarization states of the two gammas.
Figure~\ref{fig:fig4} illustrates the double Compton polarimeter configuration used to analyze the polarization correlations between the two \SI{511}{keV} gammas.

\subsection{Entanglement Witness}

An observable Entanglement Witness, ($R$), can be constructed from the measured polarization correlations, with representative values summarized in Table~\ref{tab:Entanglment_witness} (see Ref.~\cite{doi:10.1126/sciadv.ads3046} for the definition of $R$). A violation of the classical bounds provides direct evidence of quantum entanglement. Recent experiments have measured polarization correlations in positron-annihilation photons~\cite{doi:10.1126/sciadv.ads3046,Ivashkin:2022qdr,PhysRevLett.133.132502}, establishing a practical method for quantifying entanglement in the gamma-ray regime.

\begin{table}[!tbh]
\caption{Entanglement Witness values for different initial quantum states of the two gamma photons.}
\centering
\begin{tabular}{l|l}
\hline \hline
Two-gamma Initial State & Entanglement Witness \\ \hline
Uncorrelated, $\ket{H}$ or $\ket{V}$ & $R=1$ \\
Separable (not entangled, classical), 
$\ket{HV}$ or $\ket{VH}$ & $R=1.63$ \\
Einstein-Podolsky-Rosen, $ R = f(A_{p,1}, \, A_{p,2}, \, \text{hidden variables})$ & $R \leq 2$ \\
Entangled,  
$\frac{1}{\sqrt{2}} \left( \ket{HV} + \ket{VH} \right)$ & $R_{\text{max}}=2.85$ \\
\hline \hline
\end{tabular}
\label{tab:Entanglment_witness}
\end{table}

\section{Advantages of Accelerator-Based Entangled Gamma Sources}

The proposed system offers several key advantages over traditional radioactive positron sources. 
First, the accelerator-driven positron beam enables photon production rates that are several orders of magnitude higher than those achievable with conventional radioisotope sources, resulting in significantly improved statistics and reduced data acquisition times. 
Second, beam parameters, including energy, polarization, intensity, and timing structure, can be precisely controlled and optimized for specific experimental objectives. 
Third, the well-defined time structure of the accelerator beam facilitates high-precision coincidence measurements and background suppression techniques. 
Fourth, \SI{511}{keV} annihilation photons possess substantial penetrating power, enabling studies of dense materials and large-scale objects that are inaccessible to lower-energy photon sources. 
Finally, the availability of polarized positron beams introduces a unique capability for systematic investigations of quantum entanglement and spin-dependent phenomena.

An additional advantage of the accelerator-based approach is the ability to reverse the positron beam polarization in a controlled manner. 
Since the spin state of the annihilating positron influences the quantum state of the emitted photon pair, polarization reversal provides a powerful experimental handle for investigating spin correlations, testing fundamental symmetries, and studying the dependence of entanglement observables on the initial spin configuration. 

\section{Applications}

\subsection{Medical Imaging (PET Enhancement)}

Quantum-entangled annihilation gamma photons have the potential to enhance positron emission tomography (PET) by exploiting quantum correlations between the two \SI{511}{keV} photons. 
These correlations can improve coincidence selection, suppress random background, and potentially increase image contrast and spatial resolution~\cite{Watts:2020mtg,doi:10.1126/sciadv.ads3046,RevModPhys.95.021002}.

The proposed accelerator-based source is not intended for direct medical use or patient irradiation. Instead, it provides a unique high-intensity research platform for the development and optimization of next-generation entanglement-capable PET systems. 
Future PET devices will continue to rely on conventional radioactive isotopes, whose relatively low activity limits the precision and speed of detector characterization studies. 
In contrast, the proposed source can generate entangled gamma pairs at rates many orders of magnitude higher, enabling rapid and statistically precise measurements.
Such a source can be used to validate entanglement signatures, measure properties of entangled gamma rays, optimize Compton polarimeters, and evaluate coincidence-selection algorithms under controlled conditions. 
The high event rates allow these studies to be performed with unprecedented statistical accuracy and significantly reduced acquisition times. 
Furthermore, the degree of entanglement of annihilation photons depends on the annihilation environment and material properties. 
The proposed source can therefore be used to characterize tissue-equivalent materials and other samples relevant to PET imaging, creating a database of entanglement-related observables that can improve the calibration, efficiency, and reliability of future quantum-enhanced PET systems. 
The well-defined time structure of the positron beam also enables advanced timing and coincidence techniques that can isolate specific physical processes and improve signal-to-noise performance. 
In general, the proposed source provides a powerful testbed for demonstrating and advancing the practical application of quantum entanglement in medical imaging.

\subsection{Quantum Imaging and Ghost Imaging}

Entangled gamma photons enable the extension of quantum imaging techniques to the high-energy regime~\cite{325251662fcd4d999a1d597579d67827}. In ghost imaging, one photon of an entangled pair interacts with an object, while its partner is detected in a separate detector. 
Image reconstruction is achieved through analysis of the correlations between the detected photon pairs rather than direct imaging of the object itself. Such techniques may enable imaging through highly scattering, opaque, or dense materials where conventional methods are limited. 
The high penetration power of \SI{511}{keV} photons combined with quantum correlations opens new possibilities for non-destructive testing and security screening.

\subsection{Spintronics and Surface Physics} 

Positronium formation is governed by the relative spin orientations of the positron and electron. Parallel spins produce ortho-positronium, while antiparallel spins can form ortho- and para-positronium. 
By controlling positron polarization and energy, spin-dependent electron densities can be probed at surfaces, providing a sensitive tool for the characterization of spintronic and magnetic materials~\cite{PhysRevLett.126.186401}. 
Furthermore, the annihilation photons emitted from positronium decay exhibit quantum-entangled polarization states, enabling studies of spin correlations and quantum coherence in condensed matter systems. 
The combination of spin-polarized positronium spectroscopy and entangled gamma-photon detection offers a unique approach to investigate electronic surface states, magnetic ordering, and quantum phenomena at material interfaces.

\subsection{Quantum Information Science}

Although gamma photons are not presently used for practical quantum communication, entangled annihilation photons provide a unique system for exploring quantum information concepts in a previously inaccessible energy regime. 
The source can serve as a testbed for studies of entanglement generation, transport, preservation, and measurement under extreme conditions. Experiments may investigate quantum state tomography, entanglement witnesses, decoherence mechanisms, and quantum-enhanced sensing techniques. 
The ability to generate high rates of entangled gamma-photon pairs also opens opportunities for developing novel quantum metrology applications and studying quantum information processes in nuclear and high-energy environments.

\section{Summary and Outlook}

We have presented a new accelerator-driven source of quantum-entangled annihilation gamma photons. The system provides high photon intensities, precise control of beam properties, and new opportunities for entanglement studies. Future work will focus on source optimization, Compton polarimeter development, and exploration of scientific and technological applications.

\section{Acknowledgments}

This material is based upon work supported by the U.S. Department of Energy, Office of Science, Office of Nuclear Physics under Contract No. 89243126CSC000213.



\bibliographystyle{JHEP_Eric}
\bibliography{LEEPP_suleiman}

\end{document}